\documentstyle[epsfig]{article}

\newcommand{\mus}{\mbox{$\mu$s}}

\newcommand{\us}{\mbox{$\mu$s}}

\newcommand{\C}{\mbox{$^{12}$C}}

\newcommand{\N}{\mbox{$^{12}$N}}

\newcommand{\Ngs}{\mbox{$^{12}$N$_{\rm g.s.}$}}

\newcommand{\numu}{\mbox{$\nu_{\mu}$}}
\newcommand{\numub}{\mbox{$\bar{\nu}_{\mu}$}}
\newcommand{\nue}{\mbox{$\nu_{e}$}}

\newcommand{\nueb}{\mbox{$\bar{\nu}_{e}$}}

\newcommand{\nux}{\mbox{$\nu_{x}$}}

\newcommand{\ep}{\mbox{e$^{+}$}}

\newcommand{\el}{\mbox{e$^{-}$}}
\newcommand{\pos}{\mbox{e$^{+}$}}

\newcommand{\mum}{\mbox{$\mu^{-}$}}
\newcommand{\mup}{\mbox{$\mu^{+}$}}

\newcommand{\pim}{\mbox{$\pi^{-}$}}
\newcommand{\pip}{\mbox{$\pi^{+}$}}

\newcommand{\mupdecay}{\mbox{\mup\ $\rightarrow\:$ \pos $\!$ + \nue\ + \numub}}

\newcommand{\pipmup}{\mbox{\pip $\rightarrow\:$ \mup + \numu}}

\newcommand{\Ndecay}{\mbox{\Ngs\ $\rightarrow\:$ \C\ + \pos\ + \nue}}

\newcommand{\CCprot}{\mbox{p\,(\,\nueb\,,\,\ep\,)\,n }}
\newcommand{\excl}{\mbox{\C\,(\,\nue\,,\,\el\,)\,\N$_{\rm g.s.}$}}

\newcommand{\nuex}{\mbox{\nue $\rightarrow\,$\nux }}

\newcommand{\numunue}{\mbox{\numu $\rightarrow\,$\nue }}

\newcommand{\numubnueb}{\mbox{\numub $\rightarrow\,$\nueb }}

\newcommand{\NCL}{\mbox{90\%\,CL}}

\newcommand{\Dm}{\mbox{$\Delta m^2$}}
\newcommand{\sit}{\mbox{$\sin ^2(2\Theta )$}}
\newcommand{\eVc}{\mbox{eV$^2$/c$^4$}}
\newcommand{\Gdng}{\mbox{Gd\,(\,n,$\gamma$\,)}}

\newcommand{\GdngG}{\mbox{Gd\,(\,n,$\gamma$\,)\,Gd}}

\newcommand{\pnd}{\mbox{p\,(\,n,$\gamma$\,)\,d}}

\newcommand{\mean}[1]{\mbox{$\langle #1\rangle$}}

\newcommand{\apge}{\mbox{$\stackrel{>}{\sim}$}}
\begin{document}
\title{The Search for Neutrino Oscillations \numubnueb\ with KARMEN}
\author{K. Eitel and B. Zeitnitz for the KARMEN collaboration\cite{karmen}\\
	Institut f\"ur Kernphysik I, Forschungszentrum Karlsruhe\\
	Postfach 3640, D-76021 Karlsruhe, Germany}
\date{Proceedings Contribution to {\it Neutrino '98}}

\maketitle
   
\begin{abstract}
The neutrino experiment KARMEN is situated at the beam stop neutrino source
ISIS. It provides \numu 's, \nue 's and \numub 's in equal intensities from 
the \pip --\mup --decay at rest (DAR). The oscillation channel 
\numubnueb\ is investigated in the appearance mode with a 56\,t liquid 
scintillation calorimeter at a mean distance of $17.7$\,m from the 
$\nu$--source looking for \CCprot\ reactions. The cosmic induced background for
this oscillation search could be reduced by a factor of 40 due to an additional
veto counter installed in 1996. In the data collected through 1997 and 1998
no potential oscillation event was observed. 
Using a unified approach to small signals 
this leads to an upper 
limit for the mixing angle of $\sit < 1.3\cdot 10^{-3}$ (\NCL) at large \Dm\ .
The excluded area in (\sit ,\Dm) covers almost entirely the favored region
defined by the LSND \numubnueb\ evidence.
\end{abstract}	

\section{INTRODUCTION}

The search for neutrino oscillations and hence massive neutrinos is one
of the most fascinating fields of modern particle physics.
The {\bf K}arlsruhe {\bf R}utherford {\bf M}edium {\bf E}nergy {\bf N}eutrino 
experiment KARMEN searches for neutrino oscillations in different appearance 
(\numunue\ \cite{zeitnitz} and \numubnueb ) and
disappearance modes (\nuex\ \cite{nuex}). The physics program of KARMEN
also includes the investigation of $\nu$--nucleus interactions \cite{reinhard}
as well as the search for lepton number violating decays of pions and muons 
and the test of the V--A structure of \mup\ decay \cite{omega}.

Here, we present results of the oscillation search in the appearance channel
\numubnueb\ on the basis of data taken from February 1997 to April 1998 with
the upgraded experimental configuration (KARMEN2).
As will be shown in the following, no potential oscillation signal was 
observed. Therefore, special emphasis is given to the KARMEN2 capability of 
measuring $\nu$ induced events,
the determination of the \numubnueb\ evaluation cuts and the
identification and measurement of the background expectation.
  
\section{NEUTRINO PRODUCTION AND\\ EXPERIMENT CONFIGURATION}

The KARMEN experiment
is performed at the neutron spallation facility ISIS of the Rutherford 
Appleton Laboratory, Chilton, UK. The neutrinos are produced by stopping 
800\,MeV protons in a beam stop target of Ta-$D_{2}O$. In
addition to spallation neutrons, there is the production of charged
pions. The \pim\ are absorbed by the target 
nuclei whereas the \pip\ decay at rest. Muon neutrinos \numu\ therefore 
emerge from the decay \pipmup . The produced \mup\ are also stopped within the
massive target and decay via \mupdecay .
Because of this \pip -\mup -decay chain at rest ISIS represents 
a $\nu$-source with identical intensities for \numu , \nue\ and \numub\ emitted 
isotropically ($\Phi_{\nu}=6.37\cdot 10^{13}$\,$\nu$/s per flavor for 
a proton beam current $I_p=200$\,$\mu$A). 
There is a minor fraction of \pim\ decaying in flight (DIF) with the following 
\mum\ DAR in the target station which again is suppressed by muon capture of 
the high $Z$ material of the spallation target. This decay chain leads to 
a very small contamination of $\nueb/\nue < 6.2\cdot 10^{-4}$ \cite{bob}, which
is even further reduced by the final evaluation cuts.

\begin{figure}[hbt]
  \centerline{\epsfig{figure=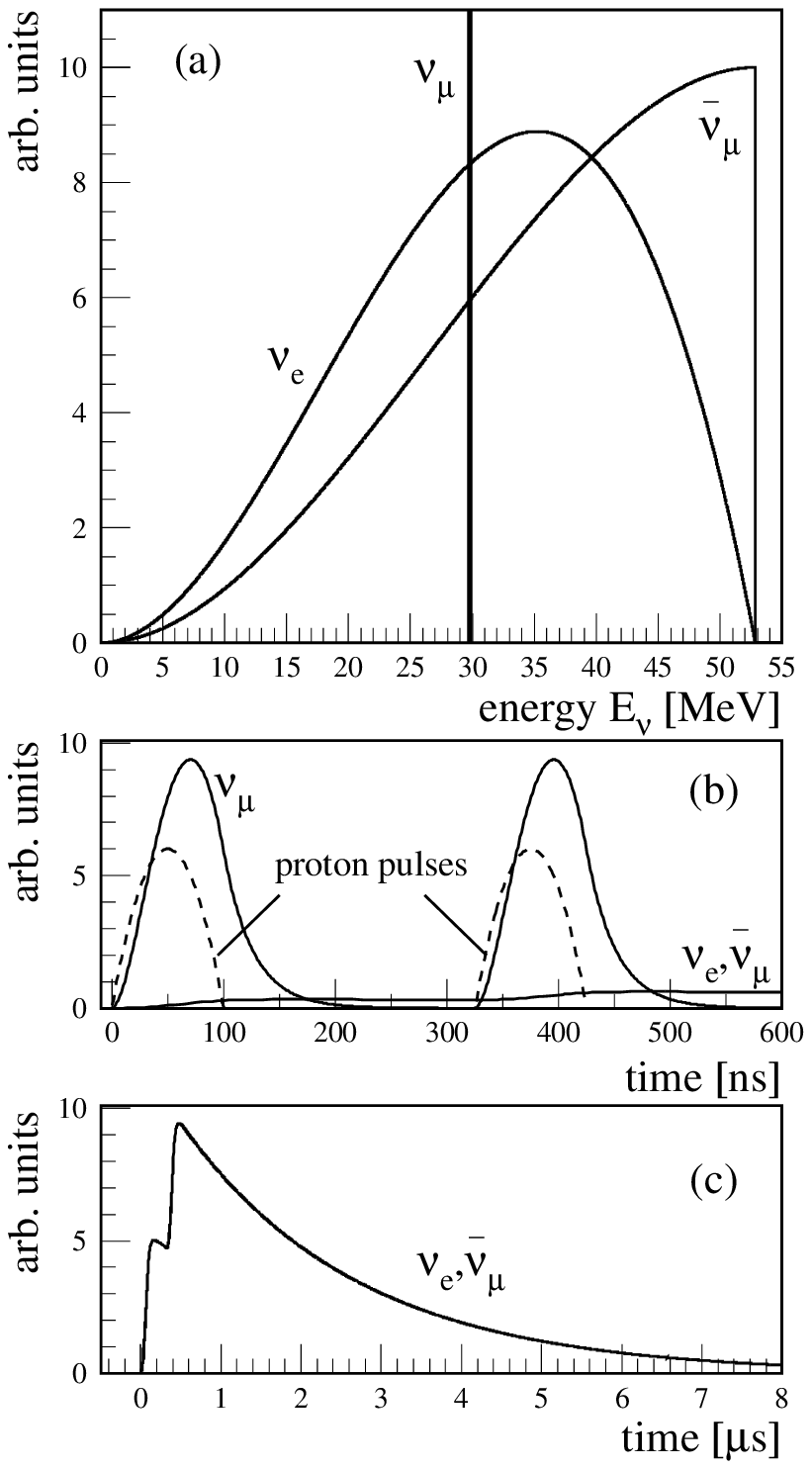,width=7.5cm}}
  \caption{Neutrino energy spectra (a) and production times of \numu\ (b)
	   and \nue ,\numub\ (c) at ISIS.}
  \label{isis_nu}
  \end{figure}

The energy spectra of the $\nu$'s are well defined due to the DAR
of both the \pip\ and \mup\ (Figure~\ref{isis_nu}a). The \numu 's from 
\pip --decay are monoenergetic with E(\numu)=29.8\,MeV, the continuous 
energy distributions of \nue\ and \numub\ up to $52.8$~MeV can be calculated 
using the V--A theory and show the typical Michel shape. 
Two parabolic proton pulses of 100\,ns base width and a gap of 225\,ns are 
produced with a repetition frequency of 50\,Hz. The different lifetimes of 
pions ($\tau$\,=\,26\,ns) and muons ($\tau$\,=\,2.2\,$\mu$s) allow a clear 
separation in time of the \numu -burst (Figure~\ref{isis_nu}b)
from the following \nue 's and \numub 's
(Figure~\ref{isis_nu}c). Furthermore the accelerator's duty cycle 
of $10^{-5}$ allows effective suppression of any beam uncorrelated background.

The neutrinos are detected in a rectangular tank filled with 56\,t of a liquid 
scintillator. This central scintillation calorimeter is
segmented by double acrylic walls with an air gap allowing efficient light 
transport via total internal reflection of the scintillation light at the 
module walls. The event position is determined by the individual module and 
the time difference of the PM signals at each end of this module. 
Due to the optimized optical properties of the organic liquid scintillator 
and an active volume of 96\% for the calorimeter, an energy resolution of 
$\sigma_E=\frac{11.5\%}{\sqrt{E [MeV]}}$ is achieved. In addition, Gd$_2$O$_3$ 
coated paper within the module walls provides efficient detection of thermal 
neutrons due to the very high capture cross section of the \Gdng\ reaction
($\sigma \approx 49000$\,barn). The KARMEN electronics is synchronized to the 
ISIS proton pulses to an accuracy of better than $\pm 2$\,ns, so that the time 
structure of the neutrinos can be exploited in full detail.

A massive blockhouse of 7000\,t of steel in combination with a system 
of two layers of active veto counters provides shielding 
against beam correlated spallation neutron background, suppression
of the hadronic component of cosmic radiation as well as reduction of the flux
of cosmic muons. On the other hand, this shielding is a source of energetic 
muon induced neutrons produced by deep inelastic muon nucleon scattering
and the nuclear capture of \mum . These neutrons produced in the steel can 
penetrate the anti counter systems undetected and simulate a \nueb\ detection 
sequence. The prompt signal is caused e.g. by a $n-p$ scattering followed by 
the delayed capture of the thermalized neutron.These neutrons were the major 
background source in the KARMEN1 experiment. In 1996 an additional third 
anti counter system with a total area of 300\,m$^2$ was installed within 
the 3\,m thick roof and the 2--3\,m thick walls of the iron shielding 
\cite{drexlin}. By detecting the muons in the steel at a distance of 1\,m 
from the main detector and vetoing the successive events this background has 
been reduced by a factor 40 compared to the KARMEN1 data.

\section{SEARCH FOR \numubnueb\ OSCILLATIONS \label{signature}}

The probability for $\nu$--oscillations \numubnueb\ can be written in a 
simplified 2 flavor description as
  \begin{equation}
	P(\numubnueb)= \sit \cdot sin^2(1.27 \frac{\Delta m^2 L}{E_{\nu}})
  \label{osdef} \end{equation}
where L and $E_{\nu}$ are given in meters and MeV, \Dm\ denotes the
difference of the squared mass eigenvalues $\Dm = |m^2_1 - m^2_2|$ in \eVc .
With $\mean{L}=17.7$\,m and $E_{\nu}<52.8$\,MeV, KARMEN is sensitive to small
mixings \sit\ for oscillation parameters $\Dm \apge 1$\,\eVc , essentially.

\begin{figure*}[hbt]
  \centerline{\epsfig{figure=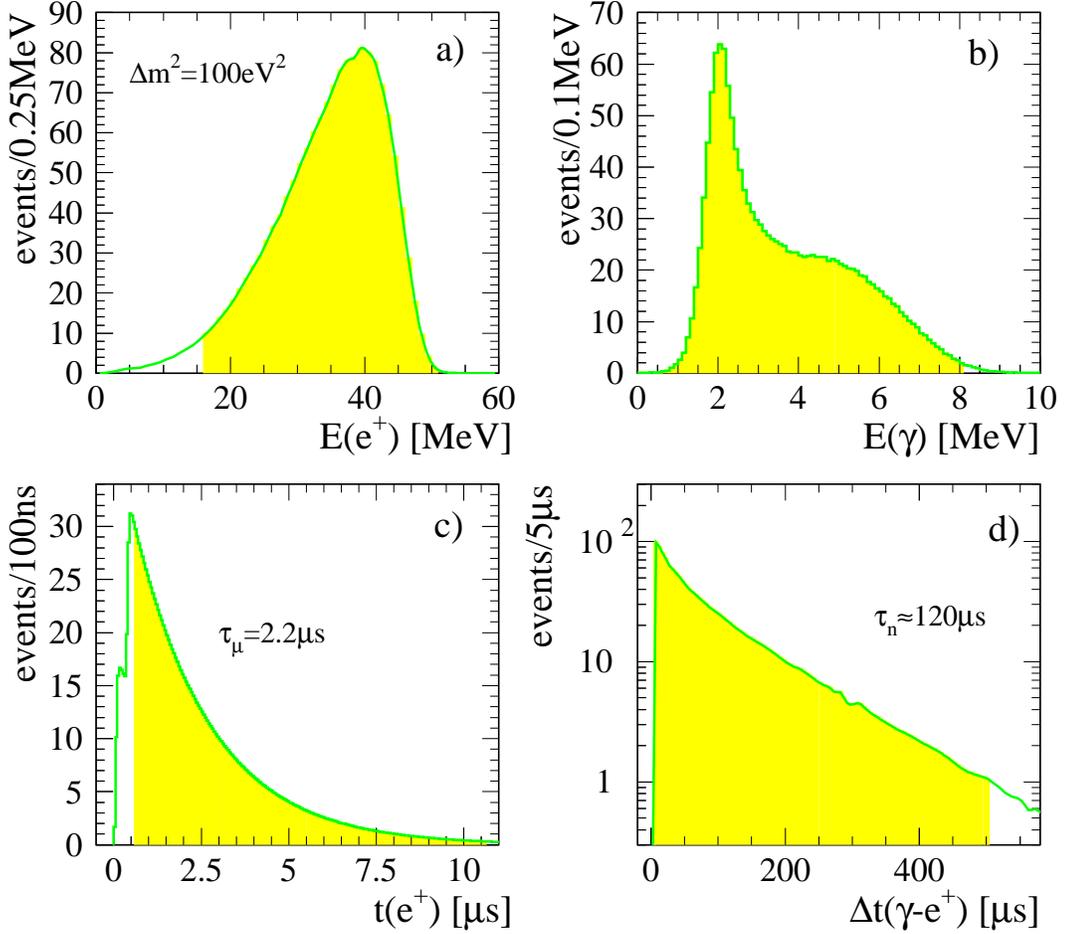,width=14.0cm}}
  \caption{Expected signature for \protect\numubnueb\ full oscillation:
	a) MC energy of prompt positron for $\Dm = 100$\,\eVc ;
	b) energy of delayed $\gamma$'s;
	c) time of prompt event relative to ISIS beam--on--target; 
	d) time difference between prompt \pos\ and delayed $\gamma$'s}
  \label{numubnueb_expect}
  \end{figure*}

The signature for the detection of \nueb 's is a spatially correlated 
delayed coincidence of positrons from \CCprot\ with energies up to 
$E_{e^+}=E_{\nueb}-Q=52.8-1.8=51.0$\,MeV (Figure~\ref{numubnueb_expect}a) and 
$\gamma$ emission of either of the two neutron capture processes \pnd\ 
with one $\gamma$ of $E(\gamma)=2.2$\,MeV or \GdngG\ with 3 $\gamma$--quanta 
in average and a sum energy of $\sum E(\gamma)=8$\,MeV
(Figure~\ref{numubnueb_expect}b).
The positrons are expected in a time window of several \mus\ after
beam--on--target (Figure~\ref{numubnueb_expect}c) with a 2.2\,\us\ exponential 
decrease due to the \mup\ decay. The time difference between the \pos\
and the capture $\gamma$ is given by the thermalization, diffusion and capture
of neutrons.

The raw data investigated for this oscillation search was recorded in the
measuring period  of February~1997 to February~1998 which corresponds to
2897\,C protons on target or $8.32\cdot 10^{20}$ \numub\ produced in the
ISIS target.  A positron candidate is accepted only if there is no activity in
the central detector, the inner anti or outer shield up to 24\,\us\ before. If
only the outermost veto counter was hit, a dead time of 14\,\us\ is applied.
These conditions reduce significantly background induced by cosmic muons:
penetrating $\mu$, decay products of stopped muons and neutrons from deep
inelastic muon scattering. Further cuts select sequences of events correlated
in space and time. To extract a possible \nueb\ induced signal these cuts were
obtained from an optimization procedure to get highest sensitivity for a
possible small (\pos ,n) signal from \numubnueb\ oscillations. This procedure
described in more detail in section~\ref{optimized_cuts} results
in the following cuts on the observed values of the prompt event: 
$0.6\;\mu\mbox{s} \le t_p \le 8.6\;\mu$s, $20.0\;\mbox{MeV} \le
E_p \le 50.0\;\mbox{MeV}$. The cuts on the delayed event are applied as
follows: $5.0\;\mu\mbox{s} \le t_s - t_p \le 270\;\mu$s, $E_s \le 7.5$~MeV and
a spatial coincidence volume of 0.97~m$^3$.

\begin{figure*}[hbt]
  \centerline{\epsfig{figure=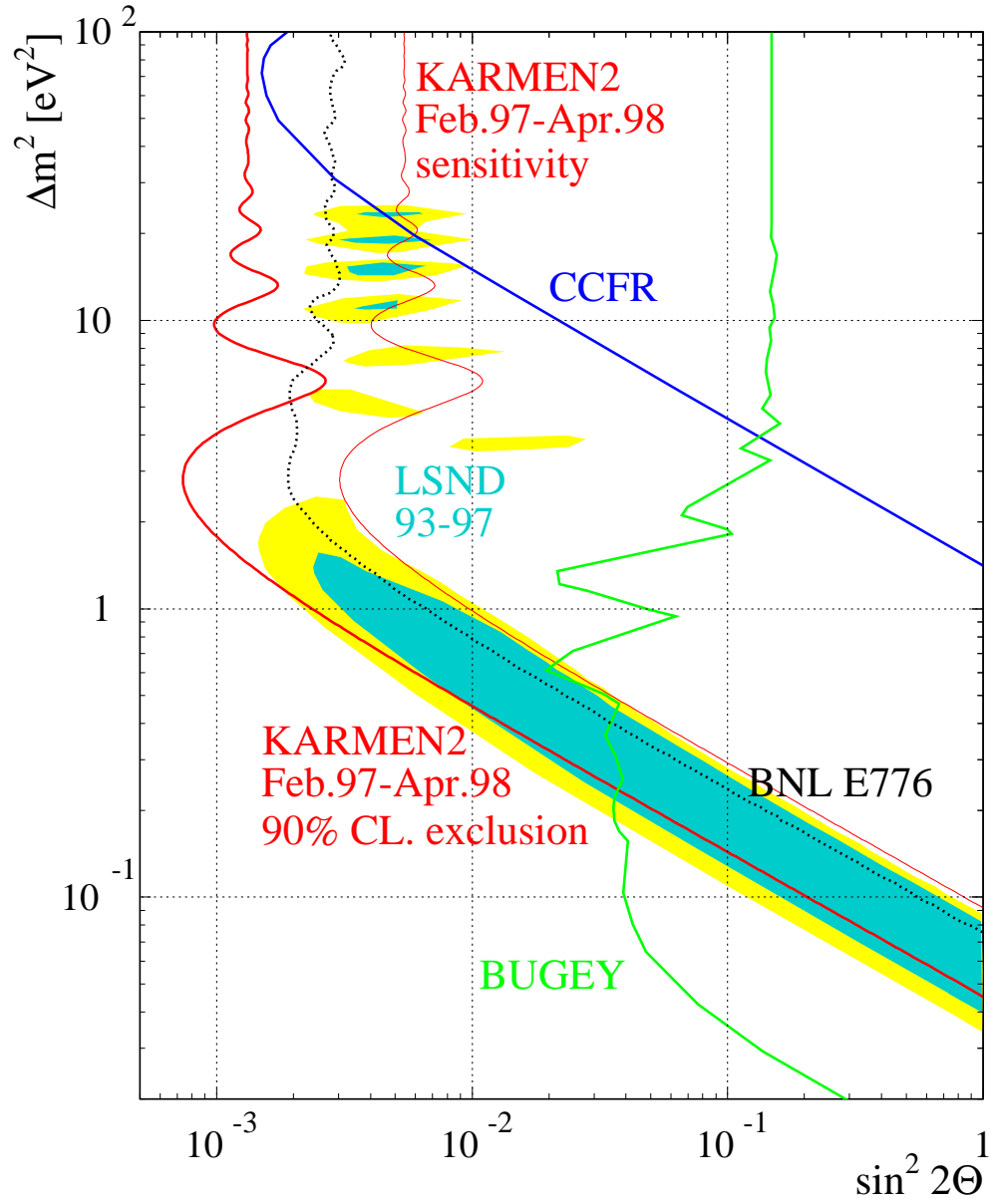,width=13.0cm}}
  \caption{KARMEN2 \NCL\ exclusion limit and sensitivity compared to
        other experiments: BNL \protect\cite{bnl}, 
	CCFR \protect\cite{ccfr}, 
	BUGEY \protect\cite{bugey} and the evidence for \numubnueb\
        oscillations reported by LSND \protect\cite{atha}. }
  \label{oszi_plot}
  \end{figure*}

In the investigated data, no sequential structure fulfilled all the required
properties for a (\pos ,n) sequence. After all cuts, the remaining 
background amounts to only $2.88\pm 0.13$ events caused by sequential cosmic 
background and $\nu$ induced sequences. These background sources are 
described in detail in the following section. The probability of 
measuring zero events with an expected number of $2.88\pm0.13$ background 
events is 5.6\%. Applying a unified approach \cite{cous}, we deduce
an upper limit of $N<1.07$ (\NCL) for a potential \numubnueb\ oscillation 
signal. With an expectation of $N=811\pm 89$ for $\sit = 1$ and large \Dm\
this corresponds to a limit of
  \begin{equation}
	\sit < 1.3 \cdot 10^{-3} \qquad (\NCL)
  \end{equation}
for $\Dm \ge 100$\,\eVc . Fig~\ref{oszi_plot} shows the KARMEN2 exclusion 
curve in comparison with other experiments.

\section{BACKGROUND SOURCES \label{background}}

There are in total four background sources contributing to the expected number
of $2.88\pm 0.13$ background events: cosmic muon induced background, \nue\
induced background dominated by the charged current reaction \excl\ with
subsequent {\em early} decays of the produced $\rm^{12}N$, random
coincidences of {\em single prong} neutrino reactions (e.g. neutral current
events) with low energy background events and the intrinsic contamination of
the neutrino source ISIS with \nueb\ from \pim\ and \mum\ decays. All
background sources except the intrinsic contamination can be measured online
and with high precision parallel to the search for neutrino oscillations.

\subsection{Cosmic muon induced background}

The KARMEN experiment is surrounded by a massive 7000\,t iron shielding.
Energetic muon induced neutrons produced in the steel can penetrate the 
anti counter systems of the detector undetected. By detecting the muons in the
steel with the new veto counter at a distance of 1\,m from the main detector 
and vetoing the successive events the background can be strongly suppressed.
This background is well described by a detailed three--dimensional Monte
Carlo simulation and therefore well understood. Moreover this background can be
precisely measured online due to the ISIS duty factor of $5\times 10^{-4}$ for
$\nu$'s from \mup\ decay. The extrapolation of the number of events recorded in
a large time window outside the beam pulses to the actual measuring time
interval of $0.6 \le t_p \le 8.6\;\mu$s is straight forward resulting in a 
precise cosmic background expectation of $0.64\pm 0.06$ events.

\subsection{\nue\ induced background}

The \numub\ for the search for \numubnueb\ are produced
by the DAR \mupdecay . Therefore there is an equal amount of \nue\
produced in the very same decay reaction. The \nue\ can be detected
via a sequence of a prompt electron from the inverse beta decay \excl\ and the
subsequent detection of a delayed positron from the decay \Ndecay . The lifetime
of \N\ is 15.9\,ms and the $\beta$--decay endpoint is 16.3\,MeV. About 500
sequences of this type have been clearly identified with a signal to noise
ratio of 35 during the KARMEN1 data taking allowing a rich analysis of the
nuclear physics involved \cite{reinhard}. 

The 1.7\% fraction of \N\ decaying within the first 270\,\mus\
contributes to the expected background level. Once again this background can be
recorded online and with high statistics. The extrapolation of the measured
number of charged current sequences to the smaller time differences and the
lower energies of delayed $\gamma$'s of the oscillation search is straight
forward. Looking for events with larger time differences ($t_s - t_p >
500$\,\mus ) and energies appropriate for ($e^+,e^-$) sequences from \excl\ and
\Ndecay\ one obtains a number of charged current events compatible with
the KARMEN1 data.

\subsection{Neutrino random coincidences}

Due to the relative high rate of low energy radioactive background events in
the KARMEN detector there is a small probability that such a low background
event occurs randomly correlated in space and time to a prompt neutrino event
with an energy above 20\,MeV. The rate of such random coincidences is strongly
suppressed by the tight spatial and time coincidence cuts. The rate of the
radioactive background events is constant in time and thus the energy and
position distributions of these events can be recorded with high precision and
extrapolated to the actual measuring time window. The probability of finding
such a low energy background event in the vicinity of a prompt neutrino event
can be obtained by generating pseudo neutrino events with a Monte Carlo method
and looking for correlated delayed events. The absolute number of random
coincidences is obtained by multiplying this probability with the measured
number of single prong neutrino reactions. Hence the amount of this background
can also be monitored online and is determined from the very same dataset
scanned for neutrino oscillations.

\subsection{Intrinsic contamination}

The only background source which can not be directly extracted from the
data is the contamination of the neutrino source with \nueb\ produced in the
\pim--\mum\ decay chain. Detailed Monte Carlo simulations \cite{bob} including
a three--dimensional model of the ISIS target are used to obtain the fraction
of \pim\ and \mum\ decaying before they are captured by the nuclei of the
target materials. The lifetimes of the \mum\ depend on the target materials and
are generally shorter than the \mup\ decay time. This effect as well as the
shape of the \nueb\ energy spectrum has been included into the calculation of
the expected number of \pos\ events generated by the contamination.

In Table~\ref{backgrd} the individual contributions of the above described
background sources are summarized. 
  \begin{table}[hbt]
  \caption{Expected sequences from different background components within
           the evaluation cuts specified in the text. The last row shows the
           expectation of (\pos ,n) sequences assuming maximal mixing, i.e.\
           $\sit = 1$.}
  \begin{tabular}{lr} \hline
        background contribution                 & events  \\ \hline
        cosmic induced sequences                & 0.64$\pm$0.06    \\
        ISIS \nueb\ contamination               & 0.56$\pm$0.09    \\
        $\nu$ induced random coincidences       & 0.72$\pm$0.04    \\
        (\el ,\ep) from \excl\ 			& 0.96$\pm$0.05    \\ \hline
        total background                        & 2.88$\pm$0.13    \\ \hline
        \nueb\ signal for $\sit=1$              &  811$\pm$89    \\ \hline
  \end{tabular}
  \label{backgrd}
  \end{table}
Note that the absolute number of expected background events is very precise.
With these reliable background contributions including the detailed knowledge
of the spectral distributions it is possible to optimize the cuts applied to 
the data without using any information about the actually measured result. 

\section{OPTIMIZATION OF EVALUATION CUTS \label{optimized_cuts}}

This section shows how we obtained optimal cuts for the search for \numubnueb\
oscillations independently from the measured result. Second, we show that
if we ignore the information of the new veto counter, which corresponds to the
KARMEN1 experimental situation, and loose the optimized evaluation cuts we
find background events in the neutrino window in good agreement with the
measured background expectation.

As the true values of the oscillation parameters \sit\ and \Dm\ are not 
known we chose to maximize the {\em sensitivity} of the experiment, i.e. we 
optimized the experiment to deliver the most stringent upper limit on \sit\ 
assuming that there are no neutrino oscillations in the sensitive \Dm\ range 
of KARMEN. Therefore the maximum sensitivity is equivalent to a minimal upper 
limit on \sit\ for a fixed \Dm.
Thus we calculated for every possible evaluation interval $I$ the ratio
$S(I,\Dm)$ of the expected number $N_{\mbox{\small expected}}$ of oscillation
events for maximal mixing $(\sit = 1)$ and the upper limit with $\alpha$
confidence level on the number of oscillation events $M_\alpha(I)$ that one
would get for the measuring interval $I$:
\begin{equation}
 S(I,\Dm) = \frac{N_{\mbox{\small expected}}}{M_\alpha(I)} 
\end{equation}
The optimal measuring interval $I$ is the one with maximal $S(I,\Dm)$. In order
to make this procedure independent of the result of the measurement the upper
limit $M_\alpha(I)$ was chosen to be the {\em mean}\/ expected upper limit
obtained by summing up all possible upper limits weighted with the
poisson probability of such a result:
\begin{equation}
 M_\alpha(I) = \sum_{n=0}^\infty
l_\alpha[n,\mu_b(I)] \cdot\frac{\mu_b^n}{n!}\;e^{-\mu_b} 
\end{equation}
Here $l_\alpha[n,\mu_b(I)]$ is the upper limit for $n$ measured events with
$\mu_b(I)$ expected background events (a zero oscillation signal was assumed).
The mean upper limit $M_\alpha(I)$ does not only depend on \Dm\ but also on the
absolute number of expected oscillation events and therefore on the measuring
time. We varied the cuts with respect to the following observables:
the energy of the prompt event $E_p$, the time of the prompt event $t_p$,
the energy of the delayed event $E_s$ and the time difference between delayed 
and prompt event $t_s - t_p$.

All cuts show a rather strong dependence on \Dm\ and change with the expected 
number of oscillation events, i.e. with increasing measuring time.
Motivated by the result of the LSND experiment \cite{atha} we chose to be most 
sensitive to \Dm\ value of $\le 0.3~\mbox{eV}^2$ resulting in the cuts
given in section~\ref{signature}.

\begin{figure*}[hbt]
  \centerline{\epsfig{figure=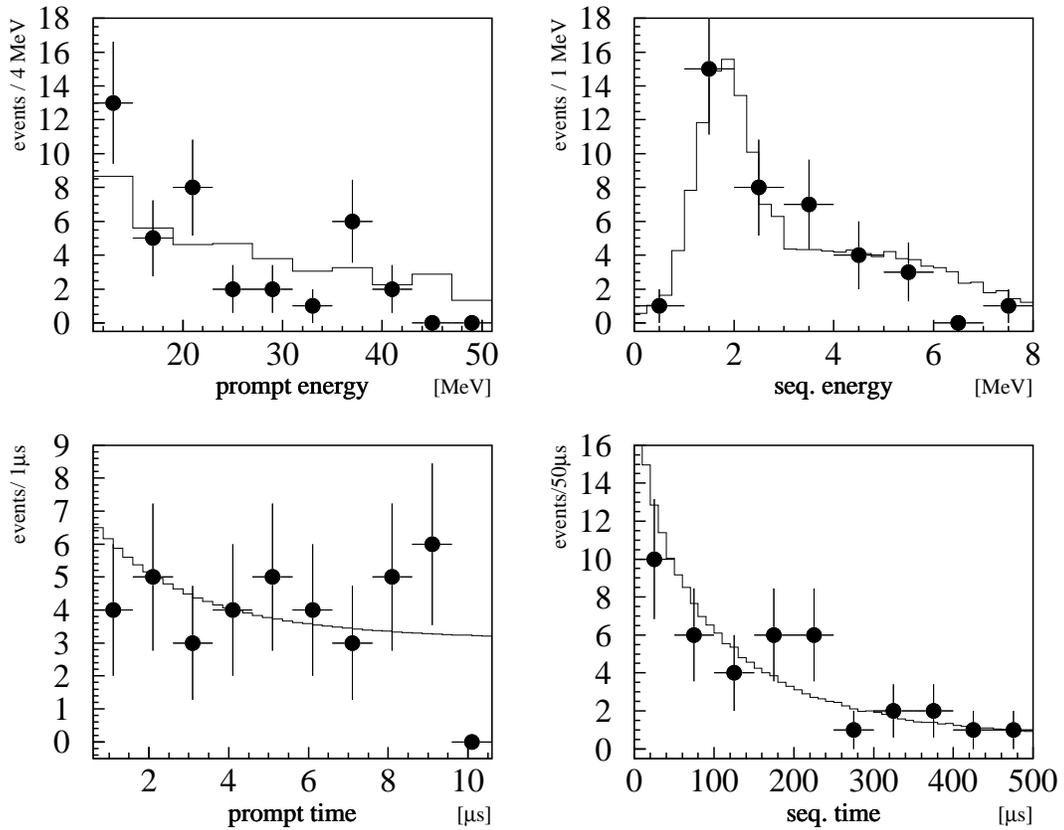,width=14.0cm}}
  \caption{Event sequences with an oscillation signature found in the KARMEN2
        data if one {\em allows} hits in the third layer veto counter.
        The number of 39 measured events agrees well with the expected number
        $40.1\pm 1.4$ events. The background is dominated by cosmic muon
        induced energetic neutrons produced in the 7000\,t steel shielding.}
  \label{noveto}
  \end{figure*}

To test the ability of KARMEN2 to measure events with a signature similar to 
that of the expected \nueb\ induced events within the appropriate $\nu$ time 
window we ignored the information provided by the additional third layer of 
veto counters. If one accepts events with an additional veto hit one obtains a
background situation similar to that of KARMEN1 dominated by cosmic ray
induced neutron background. Moreover somewhat looser cuts on the prompt energy
$11 \le E_p \le 50$~MeV, prompt time $0.6 \le t_p \le 10.6~\mu$s, energy of the
delayed event $0 \le E_s \le 8$~MeV and difference between time of delayed and
prompt event $0.5~\mu\mbox{s} \le t_s - t_p \le 500~\mu$s were used. These cuts
provide a good efficiency for muon induced neutron background. The number of
measured events in the very same dataset used for the neutrino oscillation
evaluation is 39. The expected background of $40.1\pm 1.4$ events is dominated
by $33.2\pm 1.4$ cosmic induced background sequences. Fig.~\ref{noveto} shows
energy and time distributions of the measured events compared to the expected
background. The very good agreement demonstrates once again the precise
knowledge of the background sources and the ability of KARMEN2 to detect
events with a signature similar to that of the expected oscillation events.

\section{CONCLUSION AND OUTLOOK}

We have detected no \numubnueb --like sequence in the KARMEN2 data so far. 
The knowledge of the significantly reduced background situation after the 
upgrade in KARMEN2 is precise and reliable. This is important because 
in the unified approach the upper limit for an oscillation signal depends 
on the number of expected background events even in the case of a zero result.

Our result is obtained by a frequentist approach providing full coverage which
was recently adopted by the Partice Data Group \cite{pdg}. Of course one has 
to keep in mind the true meaning of 90\% confidence intervals and that the 
obtained upper limit is due to change (it can even become less stringent) 
during the ongoing measuring time of KARMEN2. This is an unavoidable feature 
of an analysis based on such small event samples. However, the result obtained 
can indeed be used to infer physical implications concerning the result of the 
LSND collaboration which in fact shows only a 'favoured region' but not a 
detailed 90\% confidence level area.

  \end{document}